\documentclass[apj]{emulateapj}
\usepackage{apjfonts}
\submitted{2007 February 12}
\journalinfo{ACCEPTED TO ApJL}

\begin{document}
\title{THE RIBBON-LIKE HARD X-RAY EMISSION IN A SIGMOIDAL SOLAR ACTIVE REGION}
\author{CHANG LIU\altaffilmark{1}, JEONGWOO LEE\altaffilmark{1}, DALE E. GARY\altaffilmark{1}, AND HAIMIN WANG\altaffilmark{1}}
\altaffiltext{1}{Center for Solar-Terrestrial Research, New Jersey Institute of Technology, University Heights, Newark, NJ 07102-1982; chang.liu@njit.edu}

\begin{abstract}
Solar flare emissions at H$\alpha$ and EUV/UV wavelengths often appear in the form of two ribbons, which has been regarded as evidence for a typical configuration of solar magnetic reconnection. However, such a ribbon structure has rarely been observed in hard X-rays (HXRs), although it is expected as well. In this letter, we report a ribbon-like HXR source observed with the \textit{Reuven Ramaty High Energy Solar Spectroscopic Imager} (\textit{RHESSI}) at energies as high as 25--100~keV during the 2005 May 13 flare. For a qualitative understanding of this unusual HXR morphology, we also note that the source active region appeared in a conspicuous sigmoid shape before the eruption and changed to an arcade structure afterward as observed with the \textit{Transition Region and Coronal Explorer} (\textit{TRACE}) at 171~\AA. We suggest that the ribbon-like HXR structure is associated with the sigmoid-to-arcade transformation during this type of reconnection.
\end{abstract}


\section{INTRODUCTION}
The ``ribbon'' structures of solar flares have long been observed
at H$\alpha$ and EUV/UV wavelengths. A ribbon in one magnetic
polarity region is paired with a ribbon in the other magnetic
polarity region and both run parallel to the magnetic neutral line
lying between them. Such a configuration has been regarded as
evidence for the classical 2D reconnection model called the CSHKP
model \citep{carmichael64,sturrock66,hirayama74,kopp76}, in which
magnetic reconnection occurs at a coronal X-point and energy
release along the field lines produces bright flare emissions at
the two footpoints in the lower atmosphere connected to the
X-point. A series of footpoints along a coronal arcade of loops
will form two ribbons and the ribbons should separate from each
other as successive reconnections occur in the higher corona above
the arcades. Even though the actual flare process may take place
in a more complicated 3D structure, the observations of two-ribbon
flares, at least, show the general applicability of the CSHKP
model \citep{lin03}.

In many events, hard X-ray (HXR) emissions are found as single or double
compact sources near magnetic neutral lines and considered as
coming from footpoints of flaring loops \citep{ohki83,sakao96}.
Although there are found loop-top HXR sources as well
\citep{masuda95}, the trend of footpoint HXR emissions is
stronger with increasing photon energy. The footpoint emission of
HXRs is generally understood as thick target bremsstrahlung
radiation of high energy particles accelerated in the corona and
precipitating into the chromosphere along the magnetic loops
\citep{dennis88}. This means that HXRs, EUV/UV, and
H$\alpha$ commonly represent the chromospheric response to the
energy input from the corona during flares. Nevertheless, hard
X-rays sources usually appear in point-like compact regions within
the H$\alpha$/UV ribbons. This distinction between the flare
emission morphology at softer wavelengths and that of HXR
sources has been recognized as a yet unsolved problem and is a
subject of active research. One explanation was proposed by
\citet{asai02}, who found HXR kernels being confined to
stronger-field parts of the ribbons. Asai et al.'s explanation is
based on the standard magnetic reconnection model, in which the
magnetic energy release is proportional to the local field
strength. This, however, means that the confined HXR source
only represents an enhancement in energy release rate according to
the magnetic field contrast, and thus more of the ribbon in hard
X-rays should be seen, given sufficient dynamic range of the hard
X-ray observations.

In retrospect, only a single event of ribbon-like HXRs has
been reported. It was the 2000 July 14 X5.7 flare observed with
the Hard X-ray Telescope (HXT) on board \textit{Yohkoh} and the
HXR ribbons were found in both the M2- (33--53~keV) and
H-bands (53--93~keV) \citep{masuda01}. However, most of the flare
rising phase was not observed due to an HXT data gap. It was also
not shown whether those HXR ribbons coincided with ribbons
at other wavelengths.

In this Letter, we report an event where the HXR ribbon is
seen clearly during the 2B/M8.0 flare on 2005 May 13 as observed
with the \textit{Reuven Ramaty High Energy Solar Spectroscopic
Imager} \citep[\textit{RHESSI};][]{lin02}.

\begin{figure}[t]
\epsscale{1.}
\plotone{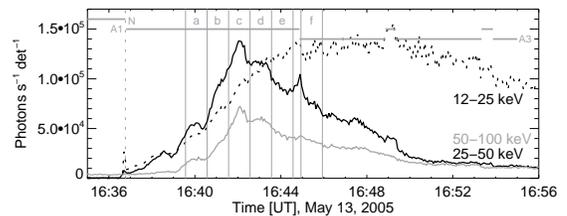}
\caption{Time profiles of \textit{RHESSI} photon rates binned into 4~s intervals. For clearer representation, the 25--50 and 50--100~keV rates are times 20 and 60, respectively. The time intervals $a$--$f$ divided by the vertical lines are for \textit{RHESSI} images shown in Figs.~2 and 3. The attenuator status for \textit{RHESSI} switched between A1 and A3 during the observation period. ``N'' denotes the time period of \textit{RHESSI} night.}
\end{figure}

\begin{figure*}[t]
\epsscale{1.}
\plotone{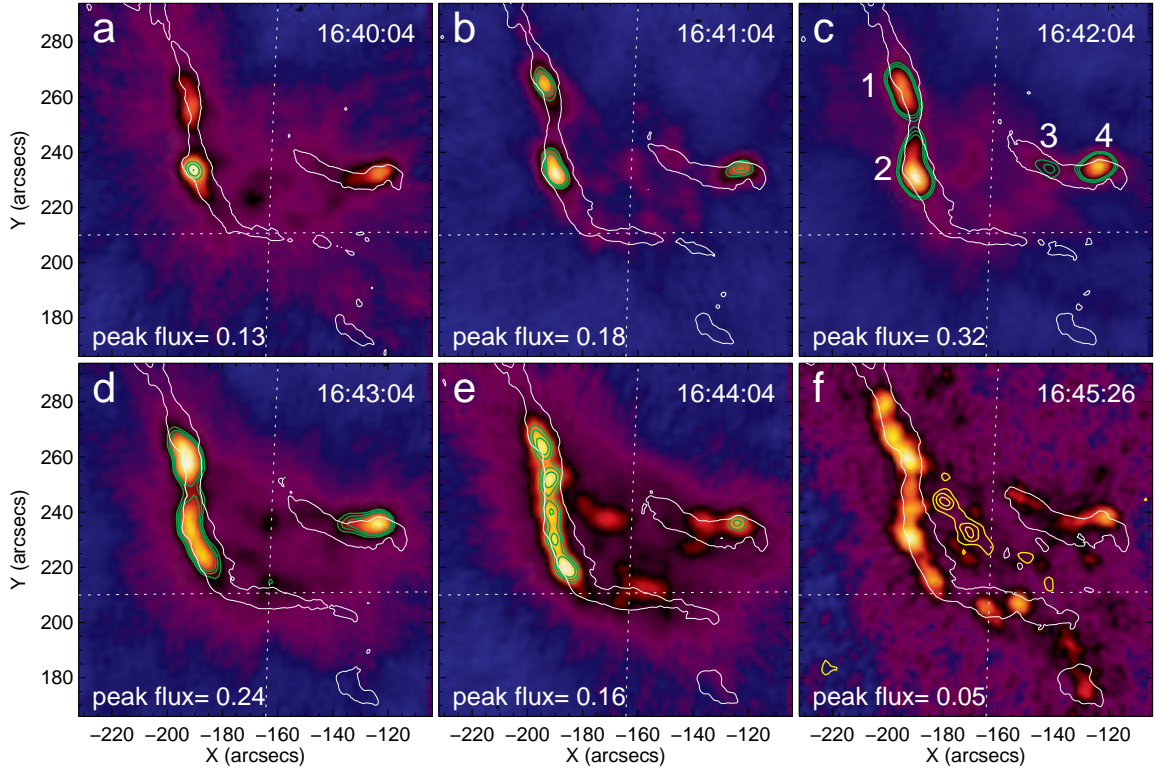}
\caption{A time sequence of \textit{RHESSI} 25--50~keV HXR images integrated in the one-minute time intervals $a$--$f$ (denoted in Fig.~1). Each \textit{RHESSI} image was reconstructed with the CLEAN algorithm using grids 1--9 with the natural weighting scheme (giving $\sim$5.9\arcsec\ FWHM resolution). The peak flux in each image is labeled and the green contours show flux at levels of 0.1, 0.115, and 0.13 photons~cm$^{-2}$~s$^{-1}$~arcsec$^{-2}$. Panel $f$ also shows \textit{RHESSI} 6--12~keV image with yellow contours at levels of 50$\%$, 70$\%$, and 90$\%$ of its maximum flux. The white contours outline the \textit{TRACE} 1600~\AA\ ribbons taken near the center of each \textit{RHESSI} time interval.}
\end{figure*}

\begin{figure*}[t]
\epsscale{1.}
\plotone{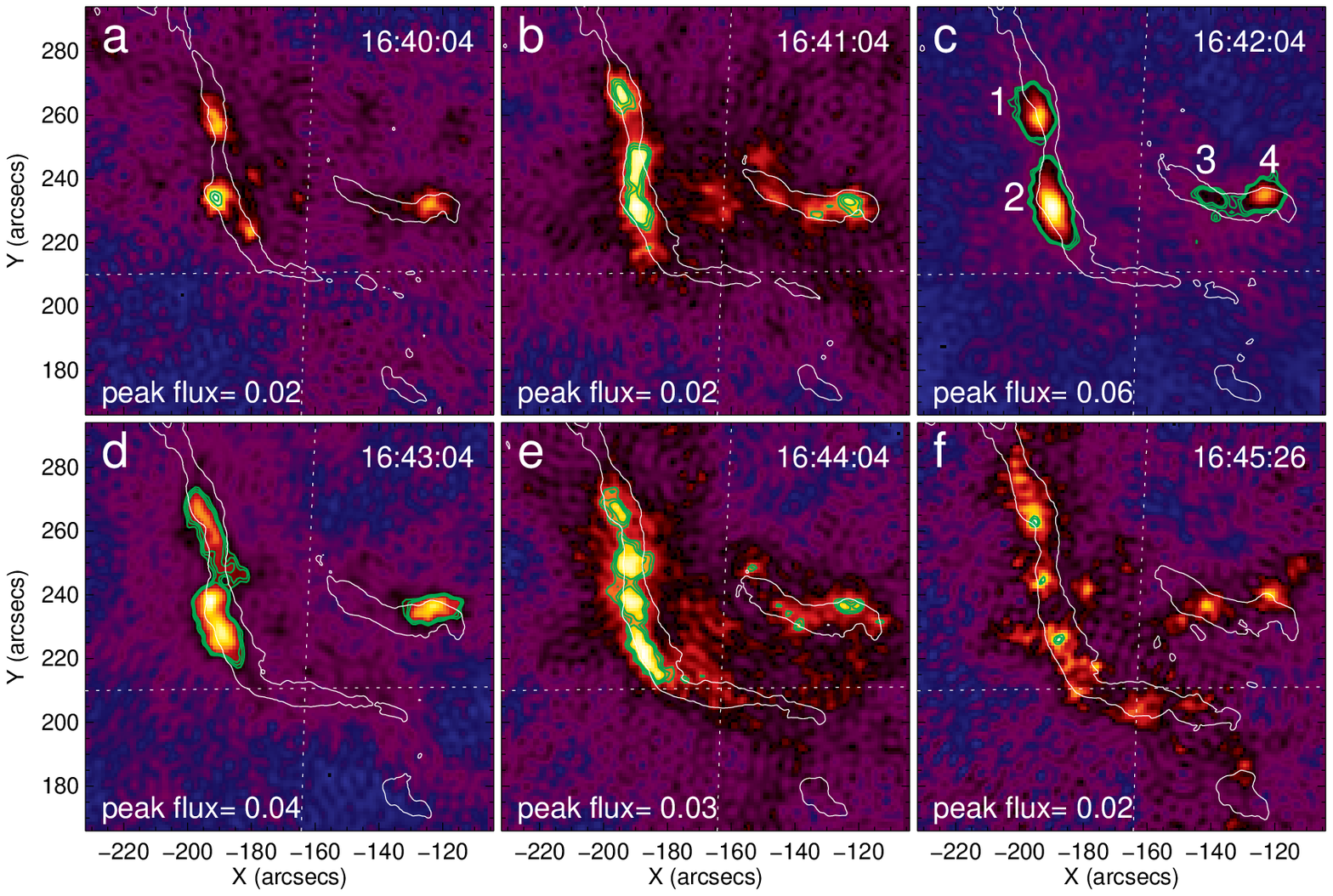}
\caption{Same as Fig.~2, but the \textit{RHESSI} 50--100~keV HXR images are shown. The green contours show flux at levels of 0.018, 0.02, and 0.022 photons~cm$^{-2}$~s$^{-1}$~arcsec$^{-2}$.}
\end{figure*}

\section{OBSERVATION}
Figure 1 shows the \textit{RHESSI} HXR lightcurves at three energy channels along with the time intervals chosen for imaging ($a$--$f$). For high image quality, we chose a one-minute time interval. We reconstructed \textit{RHESSI} images with the CLEAN algorithm using subcollimator 1--9 with natural weighting, which gives $\sim$5.9\arcsec\ FWHM resolution. The natural weighting scheme, in which counts from all detectors are given equal weight, has a better sensitivity for the detection of isolated compact sources and extended sources \cite[][also see \citealt{veronig06}]{hurford02}. We avoided imaging in the time period between intervals $e$ and $f$, within which change of the \textit{RHESSI} attenuator from A1 to A3 occurred. No pulse pile-up of lower energy photons is evident in the obtained \textit{RHESSI} HXR images. Alignment between \textit{RHESSI} and the \textit{Transition Region and Coronal Explorer} \citep[\textit{TRACE};][]{handy99} was done in two steps. First, by matching the main sunspot feature we found that the shift between \textit{TRACE} 1600~\AA\ and white-light (WL) bands for this event are negligible. We then aligned the \textit{TRACE} WL channel with an MDI intensitygram via cross-correlation and applied the offset found to the \textit{TRACE} 1600~\AA\ images. Considering that the MDI roll angle can be known no better than 1$^{\circ}$, the accuracy of alignment between \textit{RHESSI} and \textit{TRACE} 1600~\AA\ band is estimated to be $\sim$5\arcsec\ at maximum.

Figure~2 shows the \textit{RHESSI} 25--50~keV maps superposed with contours at fixed photon flux levels and those outlining \textit{TRACE} UV ribbons. Until the flare maximum (intervals $a$, $b$, and $c$), HXR emissions appear as point-like compact sources, which are located within the flare ribbons. At the flare maximum time, there are four HXR sources and the average magnetic field strengths of flare ribbons associated with HXR emission kernels ($\gtrsim 50\%$ of the maximum) are about two to three times larger than that of the other parts of the ribbons. This result is in agreement with the suggestion by \cite{asai02} that the HXR emissions concentrate on the parts of ribbons with stronger magnetic fields.

After the flare maximum (intervals $d$, $e$, and $f$), the HXR sources, however, become elongated to form a ribbon structure. This footpoint-to-ribbon evolution of HXR emissions is more evident for the much stronger eastern HXR sources. Several kernels can be seen within the ribbon during the time interval $e$. At the time interval $f$, significant HXR emission (although with a much lower flux level compared with the peak time) is found along the entire section of each ribbon. Note that the HXR distribution at this time is no longer concentrated in the strongest magnetic field regions, unlike the above mentioned suggestion by \citet{asai02}.

A nearly identical trend is found at higher energies (50--100~keV) as shown in Figure~3, although the image quality is not as good as that of 25--50~keV images due to significantly lower photon counts. At lower energies (6--12~keV), X-ray sources lie between the ribbons presumably near the tops of the loops joining them (see Fig.~2 panel $f$). Therefore this ribbon-like structure is not a phenomenon limited to low energy thermal particles, but extends to nonthermal high-energy electrons.

We also note, within the accuracy of our alignment, that the HXR sources tend to lie at the evolving edge of UV ribbons that were expanding to the southeast and northwest directions. This indicates that the HXR sources are due to the electrons precipitating along the most recently reconnected field lines. One may ask whether the HXR sources move with and/or along the UV ribbons. Although we could not trace the HXR sources in detail because of the limited time range of good count statistics for \textit{RHESSI} imaging, the location of the HXR ribbons remains consistent with the UV ribbons in all six time intervals covering 6 minutes. Image quality would have been degraded if we used shorter time intervals for imaging.

\begin{figure*}[t]
\plotone{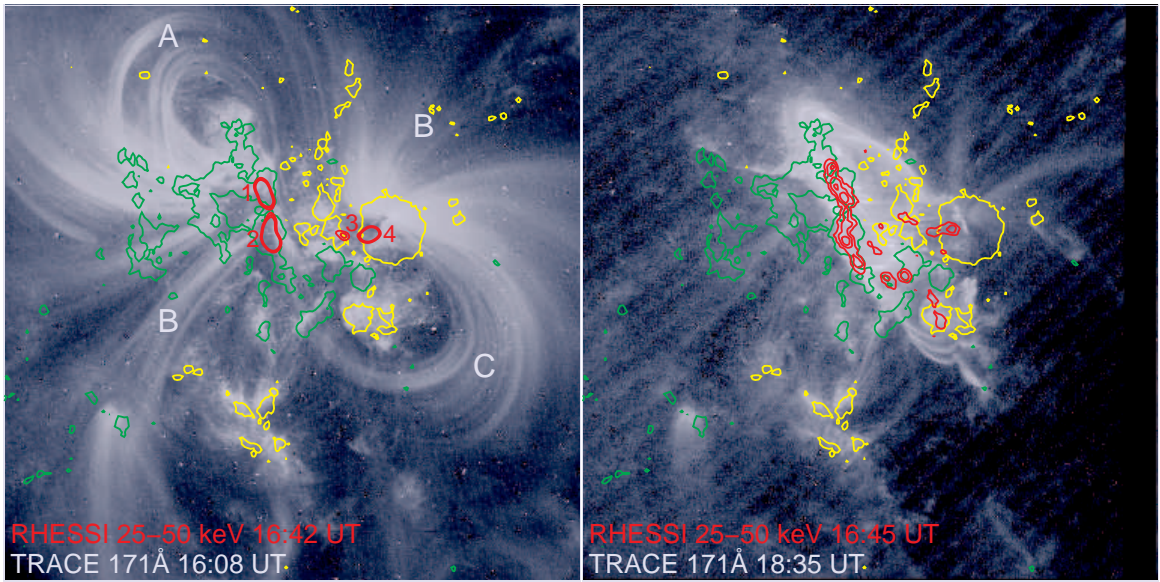}
\epsscale{1.}
\caption{Pre- and postflare images from \textit{TRACE} 171~\AA\ channel showing the sigmoid-to-arcade evolution of the coronal magnetic field. ``A'' and ``C'' denote the magnetic elbows and ``B'', envelope loops, following the nomenclature used by \cite{moore01}. \textit{SOHO} MDI longitudinal magnetic field is superimposed with the yellow and green contours representing positive and negative fields, respectively. The contour levels are $\pm$ 50~G. Same \textit{RHESSI} contours as Fig.2$c$ are overplotted onto the preflare \textit{TRACE} image. \textit{RHESSI} image Fig.~2$f$ is overplotted onto the postflare \textit{TRACE} image with contour levels 40$\%$, 60$\%$, and 80$\%$ of the maximum. The field of view is 384\arcsec\ $\times$ 384\arcsec.}
\end{figure*}

In Figure 4, we compare \textit{TRACE} 171~\AA\ images taken just before (16:08~UT) and after (18:35~UT) the event to infer the magnetic structure of the active region. The superposed \textit{yellow} and \textit{green} contours are positive and negative longitudinal fields, respectively, measured with the MDI magnetogram, showing that this is a bipolar region consisting of a main round sunspot at the leading side with the positive magnetic polarity and a trailing part of negative polarity. The preflare EUV image (\textit{left panel}) clearly shows many conspicuous sigmoidal loops at a better resolution (0.5\arcsec) than previously observed sigmoids primarily with the \textit{Yohkoh} Soft X-ray Telescope (2.45\arcsec\ per pixel in full-resolution mode). The post-eruption image (\textit{right panel}) shows loops of an arcade, thus exhibiting the sigmoid-to-arcade transformation \citep{sterling00}. Such magnetic field evolution has been explained by \citet{moore01} in terms of tether-cutting reconnection of sigmoidal fields. To follow their convention we denote the two oppositely curved magnetic elbows as ``A'' and ``C'', each of which links one polarity to the other. They loop out on opposite ends of the neutral line to form a typical sigmoid. The envelope field denoted as ``B'' extends outward, possibly overarching the sheared core field. Since this active region magnetic field structure coincides well with that of the Moore et al's model, the reconnection scenario depicted in the model may account for the magnetic field evolution of this active region as shown in Figure 4.

\section{DISCUSSION}
The ribbon-like HXR source structure presented in this
Letter is rare, having been reported only once before
\citep{masuda01}. Yet, under the simplest flare scenario,
\citep[e.g.,][]{dennis88}, such structure is expected to be
common. Two questions therefore arise: (1) why is such structure
so rare, and (2) what is unique about this event, or this
observation, that makes the structure show up so clearly? We are
unsure of the answer to the first question, but an investigation
of the second may offer some new insight.  We first examine
whether this observation is unique with regard to instrumental
sensitivity, and then discuss whether the visibility of the hard
X-ray ribbons is associated instead with a unique physical
process.

\citet{asai02} suggested that the limited sensitivity of the
instrument (specifically, \textit{Yohkoh} HXT with dynamic range
$\sim$10) may be responsible for the footpoint-like appearance of
HXR sources in general. In other word, the HXR
ribbon-like structure exists in most events, but can be seen only
with sufficiently high dynamic range. The dynamic range of
\textit{RHESSI}, as design goal, is as high as $\sim$100
\citep{hurford02}, but citations in the literature
\citep[e.g.,][]{gallagher02} give a dynamic range of only
$\sim$20 depending on individual source structure. In the present
event, the peak intensity of the ribbon source relative to the
median intensity of a background region away from the source is
$\sim$12:1 in the two maps where the ribbon structure is most
obvious (Fig. 2$f$ and 3$f$). Therefore, this observation is not
unique with respect to dynamic range. In fact, the contrast ratio
between the brightest and faintest parts of the ribbons in these
same figures is less than 6:1, indicating that the HXR ribbons
were exceptionally uniform in brightness. We conclude that
similar HXR ribbon structure, if it existed in other
events, should have been detected regardless of whether they were
observed with \textit{Yohkoh} HXT or \textit{RHESSI}. We also
note that the model of \citet{asai02}, in which the HXR brightness
is proportional to the cube of photospheric magnetic field strength, does not
fit this event since the magnetic field strength contrast along the 
ribbon measured from the MDI images is as high as $\sim$11, and does not 
show such uniformity.
Another property to note is that the HXR sources show a
typical footpoint structure in the rise phase, and only later
evolved into the ribbon-like morphology. This observation is
therefore not biased to show the ribbon structure. Nor can we
appeal to the flare strength as a unique property of this event,
since it is a moderate GOES class M8.0, and quite usual in HXR
strength. It is thus unlikely that this event's unique ribbon-like
HXR source relates to instrumental sensitivity.

We therefore regard the uniformity of brightness of the HXR
ribbons as indicative of a corresponding uniformity of energy
deposition rate by electrons accelerated in the reconnecting
current sheet between the ribbons late in the event.  To explain
the uniqueness of the event, we speculate that it may relate to
the exceptionally clear transition from sigmoid to arcade
structure seen in the TRACE images (Fig.~4). The only other
published ribbon-like HXR event, the 2000 July 14 X5.7
flare \citep{masuda01}, also involved an active region that showed
a preflare sigmoid in soft X-rays \citep{zhang05} and a postflare
arcade structure \citep{masuda01}. We acknowledge that this alone
cannot fully explain why the HXR ribbon appears only in
these two events, since many sigmoids have been observed to evolve
to magnetic arcades without the report of HXR ribbons.
Therefore, we speculate that it is not only the transformation
itself, but also the uniformity of electron acceleration along the
arcade during and after its formation that account for the
uniformity of HXR ribbon brightness. A systematic survey of the
{\it RHESSI} events in sigmoidal active regions is worthwhile, in
order to ascertain whether faint, ribbon-like HXR structure may
have been overlooked. Such a study will be needed to establish the
physical association between them, if such an association exists.


\acknowledgments
The authors thank \textit{RHESSI} and \textit{TRACE} teams for excellent data sets. We also thank the referee for helpful comments to improve the paper. CL is indebted to Brian R. Dennis for valuable discussion. CL thanks Vasyl Yurchyshyn and Na Deng for suggestions and help. This work is supported by NSF/SHINE grant ATM 05-48952. JL was supported by NSF grant AST 06-07544 and NASA grant NNG0-6GE76G.

\clearpage


\begin{thebibliography}{}
\bibitem[Asai et al.(2002)]{asai02} Asai, A., Masuda, S., Yokoyama, T., Shimojo, M., Isobe, H., Kurokawa, H., \& Shibata, K.\ 2002, \apjl, 578, L91
\bibitem[Carmichael(1964)]{carmichael64} Carmichael, H.\ 1964, in The Physics of Solar Flares, ed. W. N. Hess (NASA SP-50; Washington, DC: NASA), 451
\bibitem[Dennis(1988)]{dennis88} Dennis, B.~R.\ 1988, \solphys, 118, 49
\bibitem[Gallagher et al.(2002)]{gallagher02} Gallagher, P.~T., Dennis, B.~R., Krucker, S., Schwartz, R.~A., \& Tolbert, A.~K.\ 2002, \solphys, 210, 341
\bibitem[Handy et al.(1999)]{handy99} Handy, B.~N., et al.\ 1999, \solphys, 187, 229
\bibitem[Hirayama(1974)]{hirayama74} Hirayama, T.\ 1974, \solphys, 34, 323
\bibitem[Hurford et al.(2002)]{hurford02} Hurford, G.~J., et al.\ 2002, \solphys, 210, 61
\bibitem[Kopp \& Pneuman(1976)]{kopp76} Kopp, R.~A., \& Pneuman, G.~W.\ 1976, \solphys, 50, 85
\bibitem[Lin et al.(2003)]{lin03} Lin, J., Soon, W., \& Baliunas, S.~L.\ 2003, New Astronomy Review, 47, 53
\bibitem[Lin et al.(2002)]{lin02} Lin, R.~P., et al.\ 2002, \solphys, 210, 3
\bibitem[Masuda et al.(1995)]{masuda95} Masuda, S., Kosugi, T., Hara, H., Sakao, T., Shibata, K., \& Tsuneta, S.\ 1995, \pasj, 47, 677
\bibitem[Masuda et al.(2001)]{masuda01} Masuda, S., Kosugi, T., \& Hudson, H.~S.\ 2001, \solphys, 204, 55
\bibitem[Moore et al.(2001)]{moore01} Moore, R.~L., Sterling, A.~C., Hudson, H.~S., \& Lemen, J.~R.\ 2001, \apj, 552, 833
\bibitem[Ohki et al.(1983)]{ohki83} Ohki, K., Takakura, T., Tsuneta, S., \& Nitta, N.\ 1983, \solphys, 86, 301
\bibitem[Sakao et al.(1996)]{sakao96} Sakao, T., Kosugi, T., Masuda, S., Yaji, K., Inda-Koide, M., \& Makishima, K.\ 1996, Adv. Space Res., 17, 67
\bibitem[Sterling et al.(2000)]{sterling00} Sterling, A.~C., Hudson, H.~S., Thompson, B.~J., \& Zarro, D.~M.\ 2000, \apj, 532, 628
\bibitem[Sturrock(1966)]{sturrock66} Sturrock, P.~A.\ 1966, \nat, 221, 695
\bibitem[Veronig et al.(2006)]{veronig06} Veronig, A.~M., Karlick{\'y}, M., Vr{\v s}nak, B., Temmer, M., Magdaleni{\'c}, J., Dennis, B.~R., Otruba, W., \& P{\"o}tzi, W.\ 2006, \aap, 446, 675
\bibitem[Zhang(2005)]{zhang05} Zhang, H.\ 2005, IAU Symposium, 226, 161
\end{thebibliography}
\end{document}